\documentclass[submission]{eptcs}
\providecommand{\event}{AMMSE 2011} 

\usepackage{rascal, highlight}
\usepackage{derriclst}
\usepackage[T1]{fontenc}
\usepackage{times, mathptm} 
\usepackage[scaled=0.80]{beramono}
\usepackage{float}
\usepackage{epsfig}
\usepackage{wrapfig}
\usepackage{xspace}
\usepackage{comment}

\title{Rascal: From Algebraic Specification to Meta-Programming}
\author{Jeroen {van den Bos}${}^{1,3}$ \and Mark Hills$^{1,2}$ \and Paul Klint${}^{1,2}$ \and Tijs {van der Storm}${}^{1,2}$ \and {Jurgen J. Vinju}${}^{1,2}$
\institute{Centrum Wiskunde \& Informatica, Amsterdam, The Netherlands${}^{1}$ \\
INRIA Nord-Europe, Lille, France${}^{2}$ \\
Netherlands Forensic Institute, The Hague, The Netherlands${}^{3}$}
\email{\{Jeroen.van.den.Bos,Mark.Hills,Paul.Klint,Tijs.van.der.Storm,Jurgen.Vinju\}@cwi.nl}
}
\def\titlerunning{Rascal: From Algebraic Specification to Meta-Programming}
\def\authorrunning{J. v.d. Bos, M. Hills, P. Klint, T. v.d. Storm \& J.J. Vinju}

\def\derric{\textsc{Derric}\xspace}
\def\excavator{\textsc{Excavator}\xspace}
\def\Rascal{\textsc{Rascal}\xspace}
\def\AsfSdf{\textsc{Asf+Sdf}\xspace}
\def\Asf{\textsc{Asf}\xspace}
\def\Sdf{\textsc{Sdf}\xspace}

\def\RLSrunner{\textsc{RLS-Runner}\xspace}

\floatstyle{ruled}
\newfloat{listing}{htbp}{lol}
\floatname{listing}{Listing}

\def\Id#1{\texttt{#1}}

\begin{document} 
\maketitle

\begin{abstract}
  Algebraic specification has a long tradition in bridging the gap
  between specification and programming by making specifications
  executable.  Building on extensive experience in designing,
  implementing and using \emph{specification formalisms} that are
  based on algebraic specification and term rewriting (namely \Asf and
  \AsfSdf), we are now focusing on using the best concepts from
  algebraic specification and integrating these into a new
  \emph{programming language}: \Rascal. This language is easy to learn
  by non-experts but is also scalable to very large meta-programming
  applications.

  We explain the algebraic roots of \Rascal and its main application
  areas: software analysis, software transformation, and design and
  implementation of domain-specific languages. Some example
  applications in the domain of Model-Driven Engineering (MDE) are
  described to illustrate this.
\end{abstract}

\section{Introduction}
\label{SEC:introduction}

Algebraic specification has a long tradition in bridging the gap
between specification and
programming~\cite{Hoffmann:1982:PE:357153.357158,Goguen79}.  There has
always been a tension between algebraic specifications as mathematical
objects with certain properties and algebraic specifications as
executable objects. This tension is nicely summarized by the label
"algebraic programming".

Experience has taught us that when a formalism is made executable it
effectively becomes a programming language. Even if the language
operates on a higher level of abstraction, common engineering issues
arise when developing and maintaining specifications. Like programs,
executable specifications have bugs and thus require debugging; they
are slow and thus need to be optimized; they are complex and thus need
to be analyzed in order to be understood; and finally their life
extends beyond the first version and thus they need to be maintained to
accommodate new requirements. The nature of algebraic specification
exacerbates the difficulty of some of these common software
engineering tasks. This is due to the inherent non-deterministic
nature of algebraic specification and the complexity of highly
optimized execution platforms (term rewriters). What actually happens
at run-time, and why and when, is conceptually far removed from what
is specified.
 
\begin{figure}[t]
\begin{center}
\includegraphics[width=.7\textwidth]{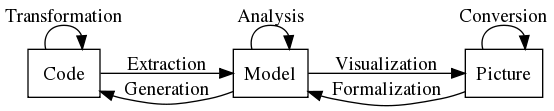}
\end{center}
\vspace{-0.6cm}
\caption{The meta-programming domain: three layers of software representation with transitions.\label{FIG:domain}}
\end{figure}
We describe the language \Rascal we are currently working on. It is a
dedicated language for meta-programming (\autoref{FIG:domain}). This means that programs can
be the input and output of \Rascal programs. \Rascal's primary
applications are in software analysis, software transformation and
design and implementation of domain-specific languages. The word
"software" should here be interpreted in a broad sense: subjects for
analysis and transformation include source code, models and meta-data
such as documentation, version histories, bug trackers, log files,
execution traces, and more. \Rascal is rooted in algebraic programming
and is targeted at solving large, real-life, problems. 

The goal of this paper is to explain why and how \Rascal is not an
algebraic specification formalism with programming language features,
but rather a programming language with algebraic specification
features. The plan of the paper is to first summarize in
Section~\ref{SEC:algebraic-perspective} the lessons we have learned in
designing and applying several languages for algebraic
programming. These lessons form the starting point for \Rascal's
design requirements. We sketch the resulting language and explain
where it deviates from the purely algebraic paradigm in
Section~\ref{SUBSECT:perspective}. In
Section~\ref{SEC:applications} we illustrate \Rascal by detailing
three applications in the domain of Model-Driven Engineering (MDE),
one of the prime application areas of \Rascal, as well as one
linking \Rascal with existing algebraic specifications. We conclude in
Section~\ref{SEC:conclusions}.

\section{An Algebraic Perspective to Meta-Programming}
\label{SEC:algebraic-perspective}

\Rascal succeeds 
\AsfSdf~\cite{BDHJJKKMOSVVV01,MetaEnv07} as our platform for
experimenting with and implementing language definitions and other
meta-programs. \AsfSdf consists of two parts: \Asf, the Algebraic
Specification Formalism (used for rewriting terms), and \Sdf, the
Syntax Definition Formalism (for specifying grammars). Below we
summarize, with hind-sight, our experiences with \AsfSdf that have
motivated the design of \Rascal. We first focus on \Asf and after that
specifically address the lessons we have learned from its combination
with \Sdf.

 
\subsection{Experience with \AsfSdf: the case of \Asf}

Our first focal point is \Asf, which originally was a standalone
formalism (\Asf,~\cite{BHK89}):
\begin{itemize}
\item An \Asf specification is modular. Modules import each other,
  while optionally instantiating sort parameters and/or renaming
  sorts.
\item Each module contains function signatures that declare (first
  order) typed functions. Functions can be either constructors (i.e.
  function names that can occur in normal forms) or defined functions
  (i.e. functions that will be eliminated by applying
  equations). Originally, the \emph{add} function on natural numbers
  (represented by the sort \texttt{NAT}) was declared in \Asf as:
  \texttt{add: NAT \# NAT -> NAT}. The latest incarnation of of
  \AsfSdf uses \Sdf~\cite{HHKR89,eelco} to define signatures
  (cf. below).
\item An equation in \Asf consists of an equality between two terms
  that respects the declared many sorted function
  signatures. Optionally, this equality may be preceded by one or more
  conditions that can be equalities or inequalities between terms. A
  conditional equality does not imply full unification: only one side of
  a positive condition may introduce variables and inequalities may not
  introduce variables at all.
\item Equations may be marked as "default" equations that apply if
  no other relevant equations apply.
\item Equations can use list matching---pattern matching modulo
  associativity of list construction---facilitating handling
  programming language constructs like statement and parameter lists.
\end{itemize}

\noindent The \emph{add} function mentioned above could be defined in
\Asf using the following equations (assuming appropriate definitions
for the \texttt{NAT} constant \texttt{0} and the successor function
\texttt{succ}):
\begin{verbatim}
[add1] add(X, 0)       = X
[add2] add(X, succ(Y)) = succ(add(X, Y))
\end{verbatim}

The design and use of \Asf have always been focused on executability
by way of (left-most innermost) term rewriting. Initial
implementations of \Asf compiled to Prolog, later ones (in the context
of \AsfSdf) to highly efficient C code. As part of \AsfSdf, \Asf has
been successfully used for the analysis and transformation of
multi-million line software systems and for the implementation of
industrial domain-specific languages~\cite{BDKKM96.ind}.  These
experiences have lead to the following observations:
\begin{itemize}
\item Although \Asf allows arbitrary rewrite rules, programmers almost
  without exception write strongly confluent and terminating sets of
  rules. They do that by introducing enough intermediate function
  symbols and by strictly using default rules when not all cases need
  to be matched by a function symbol. In other words, a locally
  non-confluent specification is almost always considered to be buggy
  rather than simply declarative.

\item Programmers in the meta-programming domain write specifications
  under the assumption of leftmost innermost reduction. By doing this
  they use \Asf as a first-order functional programming language with
  advanced pattern matching features.

\item Practically all bugs in \Asf specifications are caused by
  non-matching terms in conditions and therefore non-reducing terms
  that were supposed to be reduced. Debugging a specification amounts
  to carefully simplifying an input term to the smallest possible term
  that triggers a bug, then running the specification and locating the
  offending rewrite rule.


\item Term rewriting can be implemented extremely efficiently and
  scales to big applications in meta-programming. Much of the
  efficiency is caused by maximally sharing sub-terms which allows
  small memory footprints and equality checking in
  $O(1)$~\cite{aterms,asfcompiler}. This also implies immutability of
  data values at run-time.


\item For large-scale meta-programming, which implies signatures of
  hundreds of constructor functions for the abstract syntax trees of
  programming languages, simple recursion over deep terms needs to be
  automated. Many meta-programs are ``structure shy'': they only apply
  to some node types of the abstract syntax and such nodes may be
  buried deep in a term. We have extended \Asf with so-called
  traversal functions~\cite{vandenBrand:2003dm} to facilitate
  automatic type-safe traversal. This feature commonly reduces the
  size of an \Asf application, sometimes up to 95\% depending on the
  size of the language.

\item Rule-based programming is not for everyone: it requires special
  training and experience to use effectively. Programmers with a
  formal computer science background have no trouble using \Asf, while
  programmers without such background have difficulty adapting to this
  paradigm. They are surprised by the fact that simple things, like
  iteration over a list, may require two or more non-trivial rewrite
  rules or the use of a complicated list pattern, while other
  operations that are complex in a normal programming language are
  suddenly completely trivial. As a result, their previously acquired
  engineering skills seem useless.

\item Text-book algorithms for static analysis and program
  optimization are not easily translated into the algebraic
  paradigm. Instead one must re-think these algorithms and visualize
  their effect in run-time and memory consumption as the execution
  platform executes them. This implies that to start analyzing and
  transforming programs, one must first re-think the basics. Although
  this is interesting from an academic perspective, from the software
  engineering perspective this represents a negative investment.

\item Sets of rewrite rules and algebraic signatures are open for
  extension. One can add a new function to a certain sort and simply
  add alternatives for all the functions that process that
  sort. Example: we add a ``do-while'' construct to a language and
  then we add new rewrite rules for the extended definition of a
  control-flow graph extractor. This is called \emph{open
    extensibility}: without changing existing code functionality can
  be extended in a modular fashion.

\end{itemize}

\noindent We have used \Asf as a high-level programming language,
applying it to different forms of meta-programming. We had to extend
algebraic specification with default rules and traversal functions to
obviate the need for large amounts of boilerplate rewrite
rules. Unfortunately, as our student influx became less formally
educated, we could not keep using \Asf as a vehicle for education in
software analysis and transformation.

\subsection{Experience with \AsfSdf: the case of \Sdf}

Since the original goal of \AsfSdf was describing programming
languages, it includes a built-in facility for describing syntax: the
Syntax Definition Formalism (SDF~\cite{HHKR89}). Naturally, this
combination of parsing and rewriting makes \AsfSdf specifically apt for
the domain of meta-programming. Parsing the input source code enables
all further analysis and transformation.
The essential characteristics of \AsfSdf derived from \Sdf are:
\begin{itemize}
\item Function signatures correspond to context-free grammar rules,
  similar in semantics to EBNF. For example, the \emph{add} function
  is now declared as \texttt{NAT "+" NAT -> NAT}. Appropriate
  priorities and associativity can also be defined.
  A non-terminal is a sort, a grammar rule is a function. Furthermore,
  \Sdf supports \emph{variable} definitions, a class of non-terminals
  specifically tagged to be meta-variables.

\item In \AsfSdf equations we write concrete syntax patterns instead
  of prefix term patterns. Any production rule in a context-free
  grammar is a term constructor. The syntax of terms is completely
  user-defined.

\item \Sdf integrates lexical and context-free syntax definitions to
  generate scannerless parsers~\cite{eelco}. This helps in broadening
  the scope of programming languages that can be accepted, for example
  to allow the analysis and transformation of languages that do not
  have a separate tokenizer or to allow parsing of embedded languages
  that have different sets of reserved keywords in different contexts.

\item When modules are combined by way of import, the signatures that
  they declare are merged. In the case of \Sdf this means composition
  of complete grammars. Because only the full class of context-free
  grammars is closed under composition, \Sdf is supported by a parser
  which supports all context-free grammars, rather than a subset like
  LALR(1) or LL(1).

\item \begin{sloppypar} List matching is implemented for regular expressions in \Sdf
  notation, e.g., terms of type \texttt{Statement*}, which corresponds
  to the language of possibly empty lists of statements, may be
  matched using arbitrary list patterns.
  \end{sloppypar}

\end{itemize}

\noindent Using \Sdf as a front-end to \Asf, the \emph{add} example
can now be written as follows:
\begin{verbatim}
[add1]  X + 0       = X
[add2]  X + succ(Y) = succ(X + Y)
\end{verbatim}
Note both how the concrete syntax of the \emph{add} function (\texttt{+})
can be used in the equations without quotation, and the use of
\texttt{X} and \texttt{Y} as meta-variables ranging over the
non-terminal for recognizing natural numbers.

Work on \AsfSdf has been documented in~\cite{DHK96}. Writing and
executing \AsfSdf specifications is supported by the \AsfSdf
Meta-Environment~\cite{klint93,BDHJJKKMOSVVV01,MetaEnv07}. The following
list of observations summarize our experience in using \AsfSdf, this
time focusing on the consequences of using \Sdf:
\begin{itemize}

\item Parsing and term rewriting are the only main features of
  \AsfSdf, and they are inseparable. This is both a strength and a
  weakness. The strength is conceptual simplicity and
  expressivity. The weakness is that one must understand both together
  and this makes the learning curve steep. New users struggle to
  conceptually separate parsing from rewriting when confronted with a
  bug or unexpected behavior.

\item The upside of modular grammars is their unlimited
  composability. The downside is that no guarantee can be given
  as to whether the composed grammar is unambiguous. Solving ambiguities is
  difficult and requires expert knowledge~\cite{basten}.

\item Since a signature is defined by a context-free grammar in
  \AsfSdf, any \emph{type errors} --- providing the wrong type of
  argument to a function --- result in \emph{parse errors}. This is
  rather uninformative, and can be especially confusing to new users.

\item Some typical programming languages have non-context-free
  syntaxes. COBOL files for example have ``margins'' that are line
  based, while inside the margins the syntax is not line
  based. Consequently, it is impossible to parse such languages using
  just context-free grammars.  Users of \AsfSdf have written
  preprocessors in scripting languages such as Perl and Python to
  remove margins or indentation and put them back later.

\item The meta-programming paradigm requires ``high fidelity'' in
  rewriting source code to source code. Otherwise unimportant details
  such as whitespace and source code comments need to be retained in
  meta programs that transform existing software systems. To
  facilitate high fidelity source-to-source transformation, we have
  extended the \Asf execution engine to rewrite full parse trees
  instead of abstract syntax trees.

\item All data that is processed must first be specified as a
  context-free grammar. For example, the \AsfSdf to C compiler
  contains a grammar for a subset of ANSI C and of an intermediate
  pattern matching automaton. The COBOL control flow visualization
  tool contains a grammar of graphviz's dot formalism. The \AsfSdf
  library even contains several definitions of XML. Defining good
  grammars is hard work, but in \AsfSdf there is no way around it.

\item Program analyses frequently require the representation of
  graphs, such as control flow graphs or data dependency graphs. These
  can be easily \emph{encoded} as parse trees (which contain tree
  nodes and lists), but this representation induces a significant loss
  of efficiency. Furthermore, operations on sets, relations and graphs
  are encoded as traversals over lists, with similar loss of
  efficiency in computation.

\item What you see is what you get. Since in \AsfSdf all data is a
  parse tree, simply unparsing it renders a complete and readable
  representation of all input, output and even intermediate data
  structures. This helps in making complex algorithms debuggable.

\end{itemize}

\noindent The combination of context-free grammars and term rewriting
is very powerful, but is not without its drawbacks. All data is
required to be described by a context-free grammar, hence parsing is a
first and unavoidable step in creating any meta-programming
tool. Input, output and intermediate representations are limited by
context-free grammars: efficient representations of data that is more
complex or simpler are not available.

\subsection{Lessons learned from other formalisms}
   
The strategic programming language Stratego~\cite{stratego} was
motivated by similar experiences with \AsfSdf. Its design aims to keep
the intention of rewrite rules as algebraic equalities, but on top of
that introduces expressive rewrite strategies to compose them. In
other words, Stratego extended algebraic programming with higher-order
parameterized rule application. Stratego's strategies, which derived
from the rewriting strategies in ELAN~\cite{elan}, are a true
first-class programmable feature rather than a conservative extension
of algebraic specifications. In Stratego, the strategies drive the
computation, not the rewrite rules. We noticed that in most
meta-programming applications of Stratego the strategies rather than
the rewrite rules do the heavy lifting. This again emphasizes the
programming rather than the specification features.
 
TXL~\cite{txl} is a functional programming language intended to
implement language extensions and software transformations.  It has
surprising similarities to \AsfSdf, but does not have its roots in
algebraic specification. This is an eye opener. Although somewhat
different, pattern matching, substitution, traversal, and BNF rules
are all features of TXL, while it does not feature algebraic equations
that are applied in a non-deterministic fashion. TXL is also known
to be very successful in the area of meta-programming, so we may
hypothesize that key components of algebraic programming are success
factors in the meta-programming domain, rather than the whole
integrated concept of algebraic specification.

Among other things, from ELAN~\cite{elan} and Maude~\cite{maude} we
have learned how ACI matching (associative and commutative matching
with defined identities) can be used to express sets and relations. 
Maude is known to be
strong in analysis algorithms, such as model checking, while \AsfSdf
is naturally stronger in transformation intensive applications. We
have also experimented with a language called RScript~\cite{rscript},
inspired by relational analysis tools such as Grok~\cite{grok} and
Crocopat~\cite{crocopat}, to verify that explicit set and relation
operators would match the software analysis application domain in
combination with fact extraction implemented directly in \AsfSdf.

Another source of inspiration is ANTLR~\cite{antlr}. Although ANTLR
does not support all context-free grammars, it shines in its
applicability and popularity among meta-programmers.  This is caused
by the perspective that meta-programs need to be included in a bigger
software engineering context. ANTLR ensures by the design and
implementation of the code it generates that ANTLR-based tools can be
seamlessly integrated in ordinary software projects. A reason for this
is that the code that is generated is almost what the programmer would
write manually. ANTLR connects to the background of advanced software
engineers rather than to the background of computer scientists. This
is another eye-opener.

\section{Rationale for the Design of \Rascal}
\label{SUBSECT:perspective}

\Rascal's goal is to cover the domain of meta-programming as a whole
(see~\autoref{FIG:domain}). We now first enumerate which \AsfSdf
features are desirable to keep and which to avoid. Then we present an
overview of the features of the language.  Based on our experience,
we came to the conclusion that the following \AsfSdf features are desirable:
\begin{itemize}

\item Context-free grammars (and scannerless parsing) for the modular
  definition of the syntax of real programming languages.

\item Pattern matching (and list matching), for finding patterns in
  programs.

\item Pattern matching for dispatch over language constructs to obtain
  open extensibility.

\item Default definitions to prevent boilerplate completion of
  alternatives.

\item Automated traversal to support structure-shy applications.

\item Concrete syntax for matching and construction of source code fragments.

\item Immutability of data to facilitate efficient rewriting and for a
  safe programming environment.

\item ``what you see is what you get''. Similar to the parse trees of \AsfSdf, all data should 
have a standard, complete and human-readable serialized representation. This notation should
coincide exactly with the notation for expressions in \Rascal.

\end{itemize}

We also concluded that the following features are undesirable:
\begin{itemize}
\item Non-determinism in dispatch over language
  constructs. Meta-programs are mostly deterministic. So, simple
  rewrite rule semantics should be restricted.

\item The necessity of defining a context-free grammar for every kind
  of data. We want to re-introduce abstract data-types as separate
  feature and have the possibility to compute directly with basic
  data-types such as strings, reals and integers.

\item The paradigm of sets of rewrite rules is too exotic for many
  software engineers.

\item The all-or-nothing experience of \AsfSdf. We need a language
  that can be introduced feature-by-feature, starting from a simple
  and understandable (procedural) basis.

\item Type-checking by parsing is confusing.
\end{itemize}

We have now explained why \Rascal should be different. Now we explain
how it is different. Starting from the aforementioned features of
\AsfSdf we reorganized them into separate, independent layers and have
added features if we considered them missing. These ingredients were
then synthesized into the language design by an iterative design
process, in which we reviewed a number of key use cases. These were
static analysis algorithms, source-to-source transformations,
type-checkers and source-code generators. The resulting language is
\Rascal~\cite{KvdSV-Rascal11,rascalscam}\footnote{\url{http://www.rascal-mpl.org}}.

\Rascal is organized in a core layer which contains basic data-types
(booleans, integers, reals, source locations, date-time, lists, sets,
maps, relations), structured control flow (if, while, switch, for) and
exception handling (try, catch). To use the core you must understand
that all data is immutable and that all code is statically typed. From
this point of view, \Rascal looks like a simple general purpose
programming language with built-in, immutable data structures.

The following is a list of features that can be learned on a ``need to
know'' basis. The layers are progressively more domain specific to the
meta-programming domain:
\begin{itemize}

\item List, set, and map comprehensions for the construction of
  SQL-like queries and analyses. This includes the \texttt{<-} element
  generation operator, which can enumerate the elements of all
  container data-types, like lists, sets, maps, and trees. The same
  operator is used in for loops.

\item Algebraic data type definitions for the definition of
  (intermediate) abstract data-types. These are similar to the data
  type facilities in functional programming languages. 

\item Advanced pattern matching operators, like deep match
  (\texttt{/}), negative match (\texttt{!}), set matching and list
  matching. These can, for instance, be used in switch cases, for
  loops and comprehensions.

\item String templates with margins and an auto-indent feature. The
  margins of strings allow one to indent a template with the nesting depth
  of the \Rascal program while embedding a multi-line source code
  template. This enhances the formatting of \Rascal template-based code
  generators.

\item A \texttt{visit} statement, which is an extension of
  \texttt{switch} that traverses arbitrarily nested data in order to
  perform structure-shy analysis and transformation. Cases of a visit
  may substitute in place and/or have side effects on (local) variables by
  executing arbitrary \Rascal code. Visit is parameterized by a
  traversal strategy to allow different traversal orders.

\item A \texttt{solve} statement for fixed-point computation.

\item Syntax definitions using an EBNF-like notation for generating
  parsers. This includes disambiguation facilities.

\end{itemize}

Additionally, \Rascal is designed specifically to help the programmer
create safe, modular and generic meta-programs in the following ways: 
\begin{itemize}

\item Type inferencing for local variables in functions. Formal
  parameters and return types of functions must be explicitly
  typed. This prevents typing errors from leaking between function
  definitions.

\item There is no down-cast operator. Instead all down conversions are
  done by matching. To use a variable that is bound by a match, the
  programmer must include the match in a conditional context, such as
  an if, for, switch, visit or comprehension to ensure that in the body
  of that construct the variables are bound. As a result, \Rascal
  programs have no ``ClassCastException''-like run-time exceptions.

\item Lexically scoped backtracking. Each body of a conditional
  statement or expression that uses non-deterministic pattern
  matching may use the \texttt{fail} statement to undo the effects of
  the current scope and jump to the next available match.

\item The formal parameters of a function may also be arbitrary
  patterns, like the left-hand sides of rewrite rules. Each
  alternative for a certain function name must have mutually exclusive
  patterns. If this can not be realized, one of the
  alternatives must have the \texttt{\textbf{default}} modifier to
  indicate that it will be tried only after the other patterns
  fail. This gives us open extensibility: add a rule in a syntax
  definition or a data definition and you can add an alternative
  definition for any function that operates on that type.

\item Rascal's sub-typing lattice supports a number of layers that
  allow algorithms to work on different levels of generality. This is
  complementary to having type parameters for generic functions and
  type-parameterized abstract data-types. The \texttt{value} type is
  the top type. Algorithms that do not assume anything about a value
  use this type. The \texttt{void} type is the bottom type. This is an
  example of a longest possible sub-type chain: \texttt{void < Statement <
    Tree < node < value}. In this chain \texttt{value},
  \texttt{node} and \texttt{void} are built-in, while the others are
  defined in Rascal. The \texttt{node} type represents the common
  super-type of all abstract data-types, allowing access to and
  modification of the names and children of
  constructors. \texttt{Tree} is a library definition of an ADT for
  all parse trees and \texttt{Statement} is a defined non-terminal
  from a syntax definition for some programming language like C or
  Java. Functions may operate on each of these 5 levels, with the
  level chosen based on the amount of detail about the parameter
  that is needed. Another benefit is that the implementation of 
  parse trees, which are central to meta programming, is completely 
  transparent to the programmer.
\end{itemize}
To summarize, Rascal is a value-based procedural programming language
with high-level domain specific built-in operators. These operators
come from algebraic programming and relational calculus.  We realize
that this is not a formal definition nor a full explanation of the
language. It should, however, be a good starting point for getting an
impression of \Rascal and its design rationale.  Details of Rascal may
change as we get more feedback, and it may be extended, but this
design will not change.

\paragraph{Example}

Getting back to our running example, there are many ways in which the
\emph{add} example can be rewritten in \Rascal\footnote{This example
  is for comparison only and is a-typical since, unlike \AsfSdf,
  \Rascal has built-in, arbitrary length, integers and reals.}.  One
is to use a switch statement for case distinction:\footnote{We use the constructor \texttt{z()} to represent \texttt{0}.}

\begin{rascaldoc}NAT\SPACE{}add1(NAT\SPACE{}x,\SPACE{}NAT\SPACE{}y)\SPACE{}\{
\SPACE{}\SPACE{}\KW{switch}\SPACE{}(y)\SPACE{}\{
\SPACE{}\SPACE{}\SPACE{}\KW{case}\SPACE{}z():\SPACE{}\KW{return}\SPACE{}x;
\SPACE{}\SPACE{}\SPACE{}\KW{case}\SPACE{}succ(NAT\SPACE{}y):\SPACE{}\KW{return}\SPACE{}succ(add1(x,\SPACE{}y));
\SPACE{}\SPACE{}\}
\}
\end{rascaldoc}

\noindent This is a traditional (mostly imperative) programming style which is
fully supported. Another way of writing this same example in \Rascal is
to write a function for each case. Note that \Rascal generalizes the
notion of a function signature from a list of typed variables to a
list of patterns that may contain (possibly deeply nested) variables.
Pattern-matching at the call site determines which version of the
function is actually called (\emph{pattern-directed invocation}).  The
two functions for defining \emph{add} are then written as:

\begin{rascaldoc}NAT\SPACE{}add2(NAT\SPACE{}x,\SPACE{}z())\SPACE{}\{\SPACE{}\KW{return}\SPACE{}x;\SPACE{}\}
NAT\SPACE{}add2(NAT\SPACE{}x,\SPACE{}succ(NAT\SPACE{}y))\SPACE{}\{\SPACE{}\KW{return}\SPACE{}succ(add2(x\SPACE{},\SPACE{}y));\SPACE{}\}
\end{rascaldoc}

\noindent We can approach algebraic equations even further, since
functions that return a single expression can be abbreviated as
follows:

\begin{rascaldoc}NAT\SPACE{}add2(NAT\SPACE{}x,\SPACE{}z())\SPACE{}=\SPACE{}x;
NAT\SPACE{}add2(NAT\SPACE{}x,\SPACE{}succ(NAT\SPACE{}y))\SPACE{}=\SPACE{}succ(add2(x\SPACE{},\SPACE{}y));
\end{rascaldoc}

\noindent This example illustrates that one can stay close to algebraic
specifications, but that mixtures of algebraic style and imperative
style are supported as well.  This is very convenient when mixing, for
instance, axiom-based simplification rules with more imperative symbol
table handling.

\Rascal provides lists (with associative matching) and sets (with
associative and commutative matching) further strengthening the
algebraic flavor.  Although \Rascal remains true to its algebraic
roots, the overall feeling of the language is that of a programming
language rather than a specification language. This is not only
because we opted for a Java-like notation, but also because we have
packaged concepts differently and have introduced some non-algebraic
concepts like, most notably, global and local variables,
comprehensions, and standard control flow.  A very simple example can
illustrate this. We define binary trees with integers as leaves and
composite nodes that specify the color of the node and two
subtrees. This can be defined as follows:

\begin{rascaldoc}\KW{data}\SPACE{}ColoredTree\SPACE{}=\SPACE{}leaf(\KW{int}\SPACE{}n)
\SPACE{}\SPACE{}\SPACE{}\SPACE{}\SPACE{}\SPACE{}\SPACE{}\SPACE{}\SPACE{}\SPACE{}\SPACE{}\SPACE{}\SPACE{}\SPACE{}\SPACE{}\SPACE{}\SPACE{}|\SPACE{}composite(\KW{str}\SPACE{}color,\SPACE{}ColoredTree\SPACE{}left,\SPACE{}ColoredTree\SPACE{}right);
\end{rascaldoc}

\noindent Next we want to analyze a \texttt{ColoredTree} and compute a
frequency distribution of the colors used in composite nodes. We use a
map from integers to strings to maintain the frequencies and a local
variable \irascal{counts} to maintain this map. The automatically
inferred type of \irascal{counts} is \irascal{map[str, int]}.  A visit
statement is used that traverses an arbitrary data structure, matches
the patterns for the cases to all substructures, and executes the case
when a match is found.  The statement \irascal{counts[color]?0 += 1}
increments the current frequency count for the given color if it
exists or it increments 0 otherwise.  Note how this affects the value
of the local variable \irascal{counts}. The \Rascal code is shown in
Listing~\ref{LST:color}.

\begin{listing}
\begin{rascaldoc}\KW{public}\SPACE{}\KW{map}[\KW{str},\SPACE{}\KW{int}]\SPACE{}colorDistribution(ColoredTree\SPACE{}t)\SPACE{}\{
\SPACE{}\SPACE{}\SPACE{}counts\SPACE{}=\SPACE{}();\SPACE{}\COMM{//\SPACE{}initialize\SPACE{}an\SPACE{}empty\SPACE{}map
}\SPACE{}\SPACE{}\SPACE{}\KW{visit}(t)\SPACE{}\{\SPACE{}\SPACE{}\SPACE{}\COMM{//\SPACE{}all\SPACE{}leaves\SPACE{}and\SPACE{}composite\SPACE{}nodes\SPACE{}in\SPACE{}the\SPACE{}tree
}\SPACE{}\SPACE{}\SPACE{}\SPACE{}\SPACE{}\KW{case}\SPACE{}composite(\KW{str}\SPACE{}color,\SPACE{}\_,\SPACE{}\_):
\SPACE{}\SPACE{}\SPACE{}\SPACE{}\SPACE{}\SPACE{}\SPACE{}\SPACE{}\SPACE{}\SPACE{}\SPACE{}\SPACE{}\SPACE{}\SPACE{}\SPACE{}\SPACE{}\COMM{//\SPACE{}for\SPACE{}each\SPACE{}composite\SPACE{}node:\SPACE{}increment\SPACE{}count\SPACE{}for\SPACE{}color\SPACE{}
}\SPACE{}\SPACE{}\SPACE{}\SPACE{}\SPACE{}\SPACE{}\SPACE{}\SPACE{}\SPACE{}\SPACE{}\SPACE{}\SPACE{}\SPACE{}\SPACE{}\SPACE{}\SPACE{}\COMM{//\SPACE{}(use\SPACE{}0\SPACE{}as\SPACE{}default\SPACE{}when\SPACE{}not\SPACE{}yet\SPACE{}in\SPACE{}table)
}\SPACE{}\SPACE{}\SPACE{}\SPACE{}\SPACE{}\SPACE{}\SPACE{}\SPACE{}\SPACE{}\SPACE{}\SPACE{}\SPACE{}\SPACE{}\SPACE{}\SPACE{}\SPACE{}counts[color]\SPACE{}?\SPACE{}0\SPACE{}+=\SPACE{}1;
\SPACE{}\SPACE{}\SPACE{}\}
\SPACE{}\SPACE{}\SPACE{}\KW{return}\SPACE{}counts;
\}
\end{rascaldoc}
\caption{Counting frequencies of colors in a ColoredTree\label{LST:color}}
\end{listing}

\section{Applications}
\label{SEC:applications}

We describe three realistic applications of \Rascal to model-driven
software development here and one example of connecting \Rascal to
an existing executable specification.
\begin{itemize}
\item A DSL for Entity modeling (Section~\ref{SEC:entities}). This educational example
         is based on our submission to the 
         Language Workbench  Competition 
         2011\footnote{\url{http://www.languageworkbenches.net}}
         and illustrates
        modularity, syntax definition, AST types and code generation.

\item ECore~\cite{ECore} (Section~\ref{SEC:ecore}) is a well-known class-based
  meta-model used in many Eclipse-based modeling
  tools such as Kermeta, ATL and XText.
This example shows an
  encoding of ECore using \Rascal types, in particular the use of
  relations for DAG-like and cyclic structures.
  
\item \derric (Section~\ref{SEC:derric}) is a real-world DSL for describing
  binary file formats and is used in the digital forensics domain
  to generate data recovery tools. Despite the small size of their
  implementation, the \derric-based tools are comparable in
  functionality and performance to their industrial-strength
  counterparts currently used in practice.

\item \RLSrunner (Section~\ref{SEC:RLS}) is a library plug-in for \Rascal that enables the 
execution of existing Maude specifications using a combination of 
higher-order functions and a co-routine implemented using pipes. 
This enables us to reuse existing program analysis specifications 
instead of requiring them to be rewritten in \Rascal.
\end{itemize}

\subsection{A simple DSL: Entities} \label{SEC:entities}

\subsubsection{Concrete and abstract syntax}

\begin{listing}
\begin{rascaldoc}\KW{import}\SPACE{}lang::entities::syntax::Layout;
\KW{import}\SPACE{}lang::entities::syntax::Ident;
\KW{import}\SPACE{}lang::entities::syntax::Types;

\KW{start}\SPACE{}\KW{syntax}\SPACE{}Entities\SPACE{}=\SPACE{}entities:\SPACE{}Entity*\SPACE{}entities;
\KW{syntax}\SPACE{}Entity\SPACE{}=\SPACE{}entity:\SPACE{}\STR{"entity"}\SPACE{}Name\SPACE{}name\SPACE{}\STR{"\{"}\SPACE{}Field*\SPACE{}\STR{"\}"};
\KW{syntax}\SPACE{}Field\SPACE{}=\SPACE{}field:\SPACE{}Type\SPACE{}Ident\SPACE{}name;
\end{rascaldoc}\caption{Syntax definition of Entity models}\label{LST:entities}
\end{listing}

\noindent The Entities DSL allows you to declare entity types in order
to model business objects. An entity has named fields, which are
either primitively typed (integer, string, boolean), or contain a
reference to another entity. An excerpt of the syntax definition of
entity models is shown in Listing~\ref{LST:entities}. First, auxiliary
(syntax) modules are imported for defining Layout, Identifiers and
Types. An Entity model then consists of a sequence of zero or more
\Id{Entity}-s. An \Id{Entity} starts with the keyword \texttt{entity},
followed by a name and a sequence of zero or more
\Id{Field}s. Finally, a \Id{Field} consists of a Type (integer,
string, boolean or Entity reference) and a name. The \Id{Entities}
non-terminal is the start symbol of the Entities grammar as indicated
by the \texttt{start} keyword.


\begin{listing}
\begin{rascaldoc}\KW{data}\SPACE{}Entities\SPACE{}=\SPACE{}entities(\KW{list}[Entity]\SPACE{}entities);
\KW{data}\SPACE{}Entity\SPACE{}=\SPACE{}entity(Name\SPACE{}name,\SPACE{}\KW{list}[Field]\SPACE{}fields);
\KW{data}\SPACE{}Field\SPACE{}=\SPACE{}field(Type\SPACE{}\textbackslash{}type,\SPACE{}\KW{str}\SPACE{}name);
\KW{data}\SPACE{}Type\SPACE{}=\SPACE{}primitive(PrimitiveType\SPACE{}primitive)\SPACE{}|\SPACE{}reference(Name\SPACE{}name);
\KW{data}\SPACE{}Name\SPACE{}=\SPACE{}name(\KW{str}\SPACE{}name);
\KW{data}\SPACE{}PrimitiveType\SPACE{}=\SPACE{}string()\SPACE{}|\SPACE{}date()\SPACE{}|\SPACE{}integer()\SPACE{}|\SPACE{}boolean()\SPACE{}|\SPACE{}currency();
\end{rascaldoc}
\caption{Abstract syntax of entities}\label{LST:entitiesAST}
\end{listing}

Whereas \AsfSdf allowed only rewriting of concrete syntax trees,
\Rascal supports the automatic mapping of parse trees to ASTs, using
the library function \irascal{implode}. This function converts a parse
tree to an AST that conforms to a \Rascal ADT describing the abstract
syntax. Listing~\ref{LST:entitiesAST} shows an ADT describing the
abstract syntax of Entity models. For every syntax production in the
grammar for entities, there is a corresponding constructor in this
ADT. Every constructor has the same number of arguments as the number
of symbols in the production (modulo keywords and layout). Lexical
tokens are mapped to \Rascal primitive types.

Transformation of concrete syntax trees is useful in cases where
layout preservation is essential, such as refactoring or legacy
renovation. However, for MDE applications this is often less
important. As already discussed earlier, \Rascal addresses this issue
by providing light weight string templates, next to full-blown
source-to-source transformations.  Below we describe a simple,
template-based, Java code generator for entity models.

\subsubsection{Java code generation}

\begin{listing}
\begin{rascaldoc}\KW{public}\SPACE{}\KW{str}\SPACE{}entity2java(Entity\SPACE{}e)\SPACE{}\{
\SPACE{}\SPACE{}\KW{return}\SPACE{}\STR{"public\textvisiblespace{}class\textvisiblespace{}<}e.name.name\STR{>\textvisiblespace{}\{
\SPACE{}\SPACE{}\SPACE{}\SPACE{}\SPACE{}\SPACE{}\SPACE{}\SPACE{}\SPACE{}'<}\KW{for}\SPACE{}(f\SPACE{}<-\SPACE{}e.fields)\SPACE{}\{\STR{>
\SPACE{}\SPACE{}\SPACE{}\SPACE{}\SPACE{}\SPACE{}\SPACE{}\SPACE{}\SPACE{}'\textvisiblespace{}\textvisiblespace{}<}field2java(f)\STR{>
\SPACE{}\SPACE{}\SPACE{}\SPACE{}\SPACE{}\SPACE{}\SPACE{}\SPACE{}\SPACE{}'<}\}\STR{>
\SPACE{}\SPACE{}\SPACE{}\SPACE{}\SPACE{}\SPACE{}\SPACE{}\SPACE{}\SPACE{}'\}"};
\}
\KW{public}\SPACE{}\KW{str}\SPACE{}field2java(field(typ,\SPACE{}n))\SPACE{}\{
\SPACE{}\SPACE{}\SPACE{}\SPACE{}<t,\SPACE{}cn>\SPACE{}=\SPACE{}<type2java(typ),\SPACE{}capitalize(n)>;
\SPACE{}\SPACE{}\SPACE{}\SPACE{}\KW{return}\SPACE{}\STR{"private\textvisiblespace{}<}t\STR{>\textvisiblespace{}<}n\STR{>;
\SPACE{}\SPACE{}\SPACE{}\SPACE{}\SPACE{}\SPACE{}\SPACE{}\SPACE{}\SPACE{}\SPACE{}\SPACE{}'public\textvisiblespace{}<}t\STR{>\textvisiblespace{}get<}cn\STR{>()\textvisiblespace{}\{
\SPACE{}\SPACE{}\SPACE{}\SPACE{}\SPACE{}\SPACE{}\SPACE{}\SPACE{}\SPACE{}\SPACE{}\SPACE{}'\textvisiblespace{}\textvisiblespace{}return\textvisiblespace{}this.<}n\STR{>;
\SPACE{}\SPACE{}\SPACE{}\SPACE{}\SPACE{}\SPACE{}\SPACE{}\SPACE{}\SPACE{}\SPACE{}\SPACE{}'\}
\SPACE{}\SPACE{}\SPACE{}\SPACE{}\SPACE{}\SPACE{}\SPACE{}\SPACE{}\SPACE{}\SPACE{}\SPACE{}'public\textvisiblespace{}void\textvisiblespace{}set<}cn\STR{>(<}t\STR{>\textvisiblespace{}<}n\STR{>)\textvisiblespace{}\{
\SPACE{}\SPACE{}\SPACE{}\SPACE{}\SPACE{}\SPACE{}\SPACE{}\SPACE{}\SPACE{}\SPACE{}\SPACE{}'\textvisiblespace{}\textvisiblespace{}this.<}n\STR{>\textvisiblespace{}=\textvisiblespace{}<}n\STR{>;
\SPACE{}\SPACE{}\SPACE{}\SPACE{}\SPACE{}\SPACE{}\SPACE{}\SPACE{}\SPACE{}\SPACE{}\SPACE{}'\}"};
\}
\end{rascaldoc}\caption{Functions to generate Java source code from Entity models}\label{LST:entity2java}
\end{listing}

The code generator is shown in Listing~\ref{LST:entity2java}. It is
defined using ordinary \Rascal functions that produce string values
using \Rascal's built-in string templates.  As an example, consider
the function \Id{entity2java}. The string value returned by
\Id{entity2java} uses string interpolation in two ways. First, the
name of the \texttt{Entity e} is directly spliced into the string via
the interpolated expression \irascal{e.name.name} between \verb!<! and
\verb!>!. Next the body of the class is produced using an interpolated
for-loop. This for-loop evaluates its body (a string template again)
and concatenates the result of each iteration. For each field, the
function \Id{field2java} is called to generate a field with getter and
setter declarations. The single quote (\verb!'!) acts as margin: all
white space to its left is discarded. Furthermore, every interpolated
value is indented automatically relative to this margin. As a result
the, output of each consecutive call to \Id{field2java} is nicely
indented in the class definition. Again, for many cases, this obviates
the need for grammar-based formatters to generate readable code.

The function \Id{field2java} is an example of pattern-based dispatch,
as introduced in Section~\ref{SUBSECT:perspective}. The
\Id{field2java} function is implemented in a style reminiscent of term
rewriting. The \Id{field2java} function thus matches its first
parameter against the pattern \irascal{field(typ, n)}. This technique
is a powerful tool for implementing languages in a modular fashion. For
instance, the entity language could be extended so that entities
support computed attributes. This will involve adding new production
rules to the grammar, and new field constructors to the abstract
syntax. Finally, the code generator would have to be extended. Using
pattern-based dispatch this can be achieved by adding additional
\Id{field2java} declarations that match on the new AST constructors.
No part of the original code generator has to be modified.


\begin{figure}[t]
\begin{center}
\epsfig{file=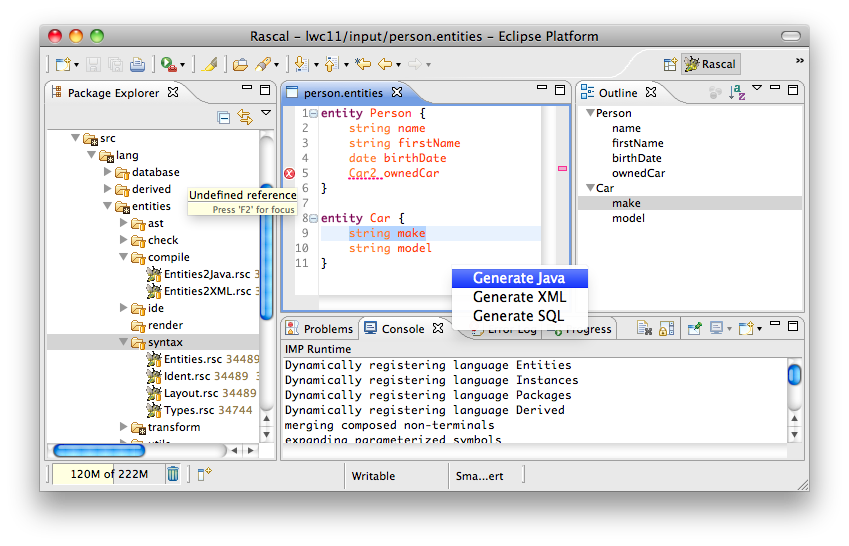,width=.95\linewidth}
\end{center}
\vspace{-1cm}\caption{Screen-shot of the dynamically generated IDE for the Entities DSL}
\label{FIG:entitiesIDE}
\end{figure}

\subsubsection{IDE support}

No language can do without IDE support, and this includes
DSLs. \Rascal exposes hooks into the Eclipse-based IMP~\cite{IMP}
framework for dynamically creating IDE support from within
\Rascal. These hooks allow the dynamic registration of, for instance, parsers, type
checkers, outliners and reference resolvers.  A screen-shot of the
generated IDE for the Entities language is shown in
Figure~\ref{FIG:entitiesIDE}. The generated IDE runs within
the \Rascal Eclipse IDE, so the package explorer on the left actually
shows the source code of the implementation of the Entities DSL. In the
middle you see an editor containing a simple entity model. It has
syntax highlighting and folding which are both based on the
context-free grammar. As you can see, there is an error: entity Person
references an undefined entity Car2.
On the right an outline is shown detailing the structure of this
entity model. Clicking on an outline element highlights the
corresponding source fragment. At the bottom of the editor
pane, (a fragment of) the context-menu is shown, including entries to
invoke various code generators.

\subsection{Relational meta-modeling: ECore} \label{SEC:ecore}
\begin{listing}
\begin{rascaldoc}\KW{data}\SPACE{}ECore\SPACE{}=\SPACE{}ecore(\KW{set}[Classifier]\SPACE{}classifiers,\SPACE{}
\SPACE{}\SPACE{}\SPACE{}\SPACE{}\SPACE{}\SPACE{}\SPACE{}\SPACE{}\SPACE{}\SPACE{}\SPACE{}\SPACE{}\SPACE{}\SPACE{}\SPACE{}\SPACE{}\SPACE{}\SPACE{}\SPACE{}\KW{rel}[Classifier,\SPACE{}Classifier]\SPACE{}subtype,\SPACE{}
\SPACE{}\SPACE{}\SPACE{}\SPACE{}\SPACE{}\SPACE{}\SPACE{}\SPACE{}\SPACE{}\SPACE{}\SPACE{}\SPACE{}\SPACE{}\SPACE{}\SPACE{}\SPACE{}\SPACE{}\SPACE{}\SPACE{}\KW{rel}[Classifier,\SPACE{}Item,\SPACE{}Type]\SPACE{}typing);
\KW{data}\SPACE{}Classifier\SPACE{}=\SPACE{}dataType(Package\SPACE{}package,\SPACE{}\KW{str}\SPACE{}name)\SPACE{}\COMM{//\SPACE{}Package\SPACE{}omitted
}\SPACE{}\SPACE{}\SPACE{}\SPACE{}\SPACE{}\SPACE{}\SPACE{}\SPACE{}\SPACE{}\SPACE{}\SPACE{}\SPACE{}\SPACE{}\SPACE{}\SPACE{}\SPACE{}|\SPACE{}class(Package\SPACE{}package,\SPACE{}Class\SPACE{}class);
\KW{data}\SPACE{}Class\SPACE{}=\SPACE{}concrete(\KW{str}\SPACE{}name,\SPACE{}\KW{list}[Item]\SPACE{}items)
\SPACE{}\SPACE{}\SPACE{}\SPACE{}\SPACE{}\SPACE{}\SPACE{}\SPACE{}\SPACE{}\SPACE{}\SPACE{}|\SPACE{}interface(\KW{str}\SPACE{}name,\SPACE{}\KW{list}[Item]\SPACE{}items)
\SPACE{}\SPACE{}\SPACE{}\SPACE{}\SPACE{}\SPACE{}\SPACE{}\SPACE{}\SPACE{}\SPACE{}\SPACE{}|\SPACE{}abstract(\KW{str}\SPACE{}name,\SPACE{}\KW{list}[Item]\SPACE{}items);
\KW{data}\SPACE{}Item\SPACE{}=\SPACE{}operation(\KW{str}\SPACE{}name,\SPACE{}\KW{set}[Option]\SPACE{}options)
\SPACE{}\SPACE{}\SPACE{}\SPACE{}\SPACE{}\SPACE{}\SPACE{}\SPACE{}\SPACE{}\SPACE{}|\SPACE{}parameter(\KW{str}\SPACE{}operation,\SPACE{}\KW{str}\SPACE{}name,\SPACE{}\KW{set}[Option]\SPACE{}options)
\SPACE{}\SPACE{}\SPACE{}\SPACE{}\SPACE{}\SPACE{}\SPACE{}\SPACE{}\SPACE{}\SPACE{}|\SPACE{}attribute(\KW{str}\SPACE{}name,\SPACE{}\KW{set}[Option]\SPACE{}options,\SPACE{}Type\SPACE{}dataType)
\SPACE{}\SPACE{}\SPACE{}\SPACE{}\SPACE{}\SPACE{}\SPACE{}\SPACE{}\SPACE{}\SPACE{}|\SPACE{}reference(\KW{str}\SPACE{}name,\SPACE{}\KW{set}[Option]\SPACE{}options);
\KW{data}\SPACE{}Type\SPACE{}=\SPACE{}classifier(Classifier\SPACE{}classifier);\SPACE{}
\end{rascaldoc}\caption{ADT for ECore (excerpt)\label{LST:ecore}}
\end{listing}

Algebraic data types generally do not support expressing
structures with sharing and/or cycles. Nevertheless, in MDE such
graph-like structures, especially class-based models, are very
common. \Rascal is a functional programming language in that all the
data is immutable. To deal with graph structures (e.g., control-flow
graphs, call-graphs, automata, work-flow models etc.) \Rascal provides
\textit{relations}. Relations are basically sets of tuples which can
be queried using comprehensions. Additionally, \Rascal provides
built-in support for computing the transitive closure of a binary
relation. 

As an example of how one might encode a class-based model as a \Rascal
data type, Listing~\ref{LST:ecore} shows an excerpt of the ECore meta
meta-model. We have used this ADT to successfully import around 300
ECore models. The top-level constructor \Id{ecore} contains a set of
\Id{Classifier}s, a subtype relation between \Id{Classifier}s and a
typing relation from \Id{Item}s (scoped in \Id{Classifier}s) to
\Id{Type}s. The last two constructor arguments capture the essential
sharing and/or cyclicity that may be present in a class model. Since
classifiers are identified by a package qualification (\Id{Package},
not shown) and a name, such values can be used as indices into the
\Id{subtype} and \Id{typing} relations.

For instance, assume we have variable \Id{class} containing a
\Id{Class} value representing a Person class. One could then find all
super classes of this class using the transitive closure of the
\Id{subtype} relation of an ECore model \Id{e}:
\begin{quote}
\begin{rascaldoc}class\SPACE{}=\SPACE{}concrete(\STR{"Person"},\SPACE{}[attribute(\STR{"name"},\SPACE{}\{\},\SPACE{}string())]);
\KW{for}\SPACE{}(sup\SPACE{}<-\SPACE{}e.subtype+[class])
\SPACE{}\SPACE{}print(\STR{"Super:\textvisiblespace{}<}sup\STR{>"});
\end{rascaldoc}\end{quote}

The encoding shown here is non-trivial and it makes a number of
trade-offs and short-cuts with respect to accurate typing of class
models. For instance, the type of the \Id{subtype} relation allows
primitive types to be sub- and super-types because they are in fact
classifiers; technically this is incorrect. Nevertheless, making the
encoding more strict would also introduce more indirections and hence,
introduce more case distinctions when processing ECore models. 

Another observation is that the encoding is very convenient for
querying, but less than optimal for transformation. Model
transformation of Ecore models encoded this way would entail creating
new \Id{ecore} values every time a single element is changed, at
arbitrary depth in the constructor. Apart from not being very
efficient, a problem with this approach is the lack of locality: every
transformation, no matter how small and localized, has to have
knowledge of the complete value. We consider this to be an area of
further research.

\subsection{A model-based approach to digital forensics: \derric} \label{SEC:derric}

Another MDE application of \Rascal is in the domain of digital
forensics. Investigations in this area are often related to recovery
of deleted, obfuscated, hidden or otherwise difficult to access
data. The software tools to recover such data require lots of
modifications to deal with different variants of file formats, file
systems, encodings etc. Additionally they are also required to return
a result within a reasonable amount of time on data sets in the
terabyte range. We are investigating a model-driven approach to this
problem by designing a DSL, \derric, to easily express the data
structures of interest. From these descriptions we generate high
performance tools for specific forensic applications.

\begin{figure}[t] 
\begin{minipage}[m]{.4\textwidth}
\begin{center}
\include{jpeg_example}
\end{center}
\vspace{-1cm}\caption{Excerpt of JPEG in \derric\label{fig:DERRIC-JPEG}}
\end{minipage}
~~~~
\begin{minipage}[t]{.55\textwidth}
\begin{minipage}[t]{\textwidth}
\begin{center}
\includegraphics[width=\textwidth]{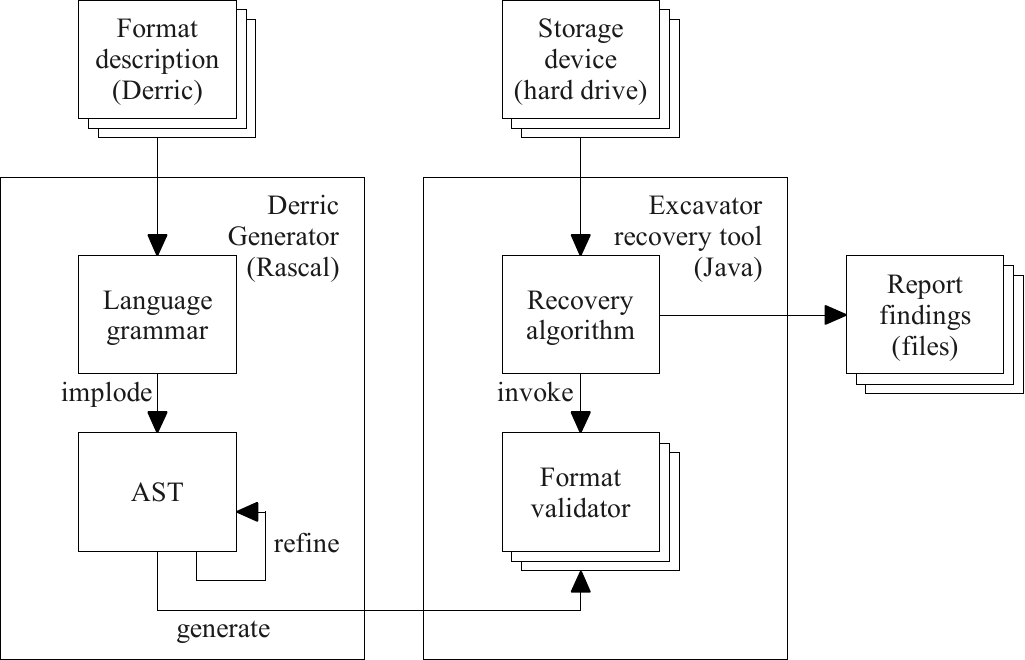}
\end{center}
\vspace{-0.5cm}
\caption{Use of \derric in \excavator\label{fig:DERRIC-PROCESS}}
\end{minipage}
\begin{minipage}[t]{\textwidth}
\begin{tabular}{|l|l|r|}\hline
Component & Lang & Size (SLOC) (total \textbf{1871})\\\hline\hline
Grammar & \Rascal & 57 \\\hline
 JPEG def & \derric & 58 \\\hline
 PNG def & \derric & 89 \\\hline
 GIF def & \derric & 101 \\\hline
 Code generator & \Rascal & 510 \\\hline
 Recovery Code & Java & 316 \\\hline
 Base library & Java & 740 \\\hline
\end{tabular}
\vspace{-0.4cm}
\caption{Sizes of the \excavator components\label{tab:DERRIC-SIZE}}
\end{minipage}
\end{minipage}
\end{figure}

\derric is a declarative data description language used to describe
complex and large-scale binary file formats, such as video codecs and
embedded memory layouts. It is essentially a very fine-grained grammar
formalism to precisely capture the way files are stored. For example,
it is possible to define a component of a file format to be a 21-bit
unsigned integer that is always stored in big endian byte
order. Figure~\ref{fig:DERRIC-JPEG} shows an excerpt of a \derric
specification for the JPEG image file format.

\derric is a language that can be used for many digital forensics
applications. Currently, we have implemented \derric in \Rascal and
have used it to develop a digital forensics data recovery tool called
\excavator (see Figure~\ref{fig:DERRIC-PROCESS}). \excavator is used
for \textit{file carving}: recovering files from storage devices
without using file system meta-data (these are often unavailable or
incomplete). \excavator is implemented as a code generator. It generates
a validator that checks whether a series of bytes conforms (or might
conform) to a certain file format.

The steps in the implementation of \excavator are shown in
Figure~\ref{fig:DERRIC-PROCESS}. The first step consists of parsing
the \derric source text and converting the parse tree to an AST
(implode). This AST is the starting point of a series of refinements
where each step takes a complete AST as input and produces a modified
AST (of the same type) that is better suited to the final goal of
generating a validator for the described format.

One refinement consists of annotating the AST with derived values
required in later stages. For instance, such values could include size
and offset information, used by compile-time expressions in the
descriptions (e.g., \texttt{lengthOf} and \texttt{offset} on lines 4
and 5 in Figure~\ref{fig:DERRIC-JPEG}).

Another refinement consists of simplifying the AST by performing
compiler optimizations such as constant folding and propagation (e.g.,
replacing string constants such as on line 6 in
Figure~\ref{fig:DERRIC-JPEG} by a list of bytes corresponding to the
string in the defined encoding, so that the generated code only has to
do simple byte comparisons).

Finally, the \derric implementation supports a number of optional
refinements, which can be executed on demand. An example is to replace
parts of the AST with alternatives that result in code that is either
faster or more precise. As such, this allows users to configure the
trade-off between accuracy and runtime performance on a case-by-case
basis.

The final step of Figure~\ref{fig:DERRIC-PROCESS} consists of
generating code. We currently have a code generator that generates
Java code and as a result, all \derric types are annotated with a
target type that maps cleanly onto Java types. For instance, a 32-bit
unsigned type will be stored in a 64-bit signed type since Java does
not support unsigned types. The resulting code is then loaded by the
\excavator runtime system to recover files from disk images.

To evaluate \excavator, we have compared it to three
industrial-strength carving tools on a set of standard
benchmarks~\cite{vdBosVdStorm11}. Our evaluation shows that even
though the implementation is very small (see
Figure~\ref{tab:DERRIC-SIZE}), it performs as good as the competing
tools both in terms of functionality and runtime performance, while
providing a much higher level of flexibility to the user.

\subsection{\Rascal front-ends for K program analysis semantics}
\label{SEC:RLS}

The K~\cite{rosu-serbanuta-2010-jlap} semantics framework is an
executable framework for defining the semantics of programming languages.
Semantics for program analysis can be defined similarly to those
for standard evaluation, with rules evaluating program constructs
over abstract value domains. Work in this area includes matching
logic~\cite{DBLP:conf/amast/RosuES10}, a logic with similar goals
to separation logic~\cite{DBLP:conf/csl/OHearnRY01}; and analysis
policies, where a policy is defined as a combination of a generic
analysis semantics for a language, specific extensions for each
specific analysis, and an annotation language tailored to each
analysis~\cite{hills-rosu-rta-2010}.

\begin{figure}
\centering
\includegraphics[width=.9\textwidth]{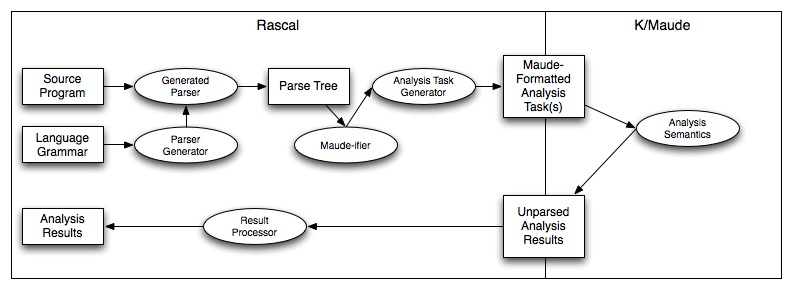}
\vspace{-2ex}
\caption{Integrating \Rascal with K Definitions in Maude}
\vspace{-3ex}
\label{fig:IntegrationModel}
\end{figure}

One limitation of this work is that it has focused on the semantics,
but not on the entire tool chain. This means that the process of
transforming the input program into something that can be evaluated
in a K semantics, or of taking the results and providing them to
the user of the analysis in some useful form, has always been 
approached in an ad-hoc fashion. While not a theoretical problem,
this makes the analyses much less useful in practice.

\begin{figure}[b!]
\vspace{-3ex}
\centering
\includegraphics[width=.7\textwidth]{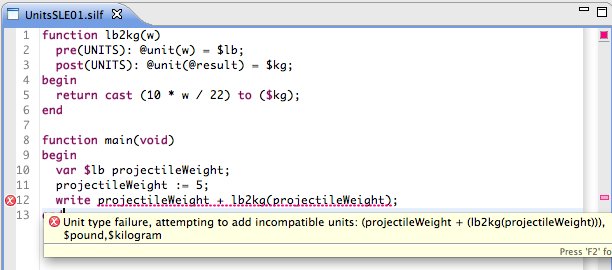}
\vspace{-2ex}
\caption{Units Arithmetic Error, Shown in Eclipse}
\label{fig:SILFUnitsProblems}
\end{figure}

The RLSRunner tool~\cite{hills-klint-vinju-sle11} provides a
solution to this for K definitions compiled to run in 
Maude~\cite{DBLP:conf/maude/2007}, a language and engine for
defining, evaluating, and reasoning about rewriting 
logic~\cite{meseguer-jtcs-1992} specifications. An overview of
the RLSRunner integration with Maude is shown in 
Figure~\ref{fig:IntegrationModel}. First, using \Rascal,
one defines the grammar for the language being analyzed, which
is used to generate a parser for the language. As shown
earlier, this automatically provides for a basic IDE for the language.
A {\it maudeifier} is then defined, allowing the parse tree
of the program to be analyzed to be transformed into a prefix form easily
consumable by Maude. This prefix form also includes location
information, encoded using a K definition for locations,
which can be used to tag errors with the location of the offending
construct. RLSRunner library functions are then used to register
handlers both for preparing the term to be given to
Maude and for parsing the term resulting from evaluation. Other
RLSRunner functions allow \Rascal to interact with Maude
and with the Eclipse environment, allowing error information
returned as a result of the analysis to be shown in the IDE and in 
the Problems view. An example of showing this information in
Eclipse is provided in Figure~\ref{fig:SILFUnitsProblems} for 
a units of measurement analysis in a simple imperative language.

\vspace*{-0.3cm}
\section{Concluding Remarks}
\label{SEC:conclusions}

We have presented the lessons we learned going from algebraic \emph{specification} languages to algebraic \emph{programming} languages. These lessons were input for the design rationale for the \Rascal language: a domain-specific programming language for meta-programming. \Rascal is easy to teach, learn and use in the domain of meta programming, but is still true to its algebraic specification roots.

\vspace*{-0.2cm}

\bibliographystyle{eptcs}
\bibliography{cited}
\end{document}